\documentclass[final]{cvpr}

\usepackage{times}
\usepackage{epsfig}
\usepackage{graphicx}
\usepackage{amsmath}
\usepackage{amssymb}
\usepackage{multirow}

\usepackage{caption}
\usepackage{subcaption}

\usepackage[pagebackref=true,breaklinks=true,colorlinks,bookmarks=false]{hyperref}

\newcommand{\tablestyle}[2]{\setlength{\tabcolsep}{#1}\renewcommand{\arraystretch}{#2}\centering\footnotesize}
\makeatletter\renewcommand\paragraph{\@startsection{paragraph}{4}{\z@}
  {.5em \@plus.2ex \@minus.2ex}{-.5em}{\normalfont\normalsize\bfseries}}\makeatother
\usepackage{amssymb,booktabs,tabulary,multirow}

\newcolumntype{x}[1]{>{\centering\arraybackslash}p{#1pt}}
\newlength\savewidth

\def\cvprPaperID{55} 
\def\confYear{CVPR 2021}

\begin{document}

\title{HINet: Half Instance Normalization Network for Image Restoration}

\author{Liangyu Chen$^1$ \thanks{Equally contribution.}\hspace{20pt} Xin Lu$^1$ \footnotemark[1] \hspace{20pt} Jie Zhang$^{1,2}$\hspace{20pt} Xiaojie Chu$^{3}$ \hspace{20pt} Chengpeng Chen$^1$\\ 
{$^1 $} MEGVII Technology \hspace{20pt} {$^2 $} Fudan University \hspace{20pt} {$^3 $} Peking University\\
{\tt\small \{chenliangyu,luxin,chenchengpeng\}@megvii.com}\\
{\tt\small j\_zhang19@fudan.edu.cn \hspace{20pt} chuxiaojie@stu.pku.edu.cn}\\
}
\maketitle

\begin{abstract}
   In this paper, we explore the role of Instance Normalization in low-level vision tasks. Specifically, we present a novel block: Half Instance Normalization Block (HIN Block), to boost the performance of image restoration networks. 
   Based on HIN Block, we design a simple and powerful multi-stage network named HINet, which consists of two subnetworks. With the help of HIN Block, HINet surpasses the state-of-the-art (SOTA) on various image restoration tasks. For image denoising, we exceed it 0.11dB and 0.28 dB in PSNR on SIDD dataset, with only 7.5\% and 30\% of its multiplier-accumulator operations (MACs), $6.8\times$ and $2.9\times$ speedup respectively. For image deblurring, we get comparable performance with 22.5\% of its MACs and $3.3\times$ speedup on REDS and GoPro datasets. For image deraining, we exceed it by 0.3 dB in PSNR on the average result of multiple datasets with $1.4\times$ speedup. With HINet, we won the 1st place on the NTIRE 2021 Image Deblurring Challenge - Track2. JPEG Artifacts, with a PSNR of 29.70. The code is available at \url{https://github.com/megvii-model/HINet}.
\end{abstract}

\vspace{-0.2cm}
\section{Introduction}
Normalization is widely used in high-level computer vision tasks: Batch Normalization~\cite{ioffe2015batch} and IBN~\cite{pan2018two} in classification~\cite{ma2018shufflenet}, Layer Normalization~\cite{ba2016layer} in DETR~\cite{carion2020end} and GroupNorm~\cite{wu2018group} in FCOS~\cite{tian2019fcos} for detection \etc. Besides, Instance Normalization~\cite{ulyanov2017improved} is used to style/domain transfer~\cite{pan2018two, huang2017arbitrary} tasks. However, the simple application of normalization to low-level computer vision problems can be suboptimal. For example, Batch Normalization can't improve the performance of the network in super-resolution~\cite{lim2017enhanced}.

\begin{figure}[t]
    \centering
    \includegraphics[width=0.47\textwidth]{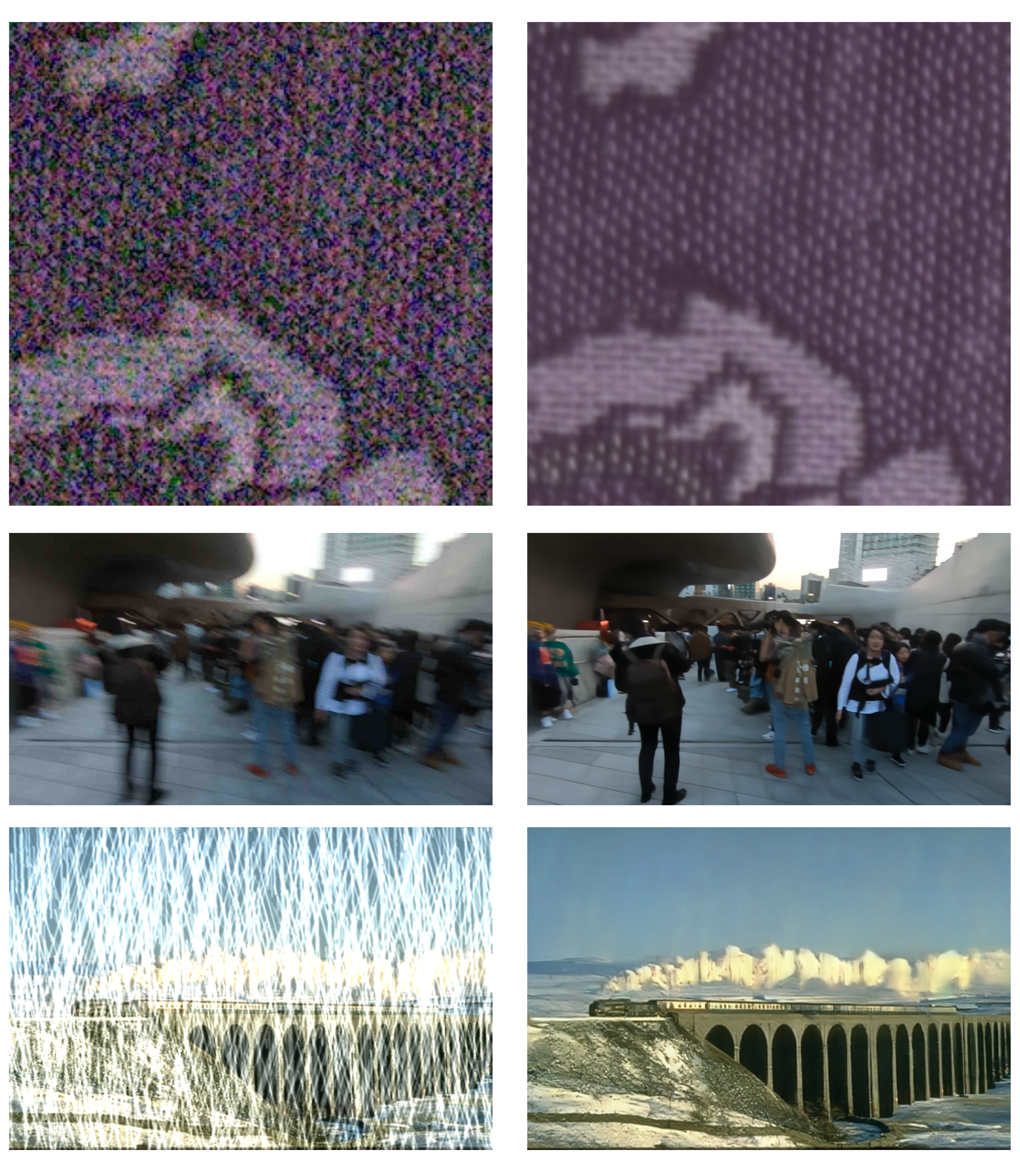}
    \vspace{-0.2cm}
    \caption{Visualized results of HINet on various image restoration tasks. Left: degraded image. Right: the predicted result of HINet. From top to bottom: image denoising, image deblurring, and image deraining task respectively.}
    \label{fig:visual_3}
    \vspace{-0.2cm}
\end{figure}

In this paper, we carefully integrate Instance Normalization as building blocks to advance the network performance in image restoration tasks. Specifically, we present a Half Instance Normalization Block (HIN Block). Based on HIN Blocks, we further propose a multi-stage network called HINet, which consists of two subnetworks. By stacking HIN Block in each subnetwork's encoder, the receptive field at each scale is expanded, and the robustness of features is also improved. In addition to the architecture of each stage, we adopt cross-stage feature fusion~\cite{zamir2021multi} and supervised attention module~\cite{zamir2021multi} between two stages to enrich the multi-scale features and facilitate achieving performance gain respectively. 


Compared with the state-of-the-art architecture MPRNet~\cite{zamir2021multi}, HINet surpasses it on various image restoration tasks. For image denoising, we exceed it 0.11 dB and 0.28 dB in PSNR on SIDD~\cite{abdelhamed2018high} dataset, with only 7.5\% and 30\% of its multiplier-accumulator operations (MACs), $6.8\times$ and $2.9\times$ speedup respectively. For image deblurring, we get comparable performance with 22.5\% of its MACs and $3.3\times$ speedup on REDS~\cite{nah2019ntire} and GoPro~\cite{nah2017deep} datasets. For image deraining, we exceed it by 0.3 dB in PSNR on the average result of multiple datasets following~\cite{Zamir2021MPRNet}, with $1.4\times$ speedup. Visualized results of various image restoration tasks are shown in Figure~\ref{fig:visual_3}. In addition, we apply HIN to various models and various datasets, the results demonstrate the generalization ability of HIN. For example, with the help of HIN, DMPHN~\cite{Zhang_2019_CVPR} increased 0.42 dB in PSNR on GoPro~\cite{nah2017deep} dataset. 


Our contributions can be summarized as follows:
\begin{itemize}
   \item We carefully integrate Instance Normalization as building blocks and proposed a Half Instance Normalization Block. To the best of our knowledge, it is the first model to adopt normalization \emph{directly} with state-of-the-art performance in image restoration tasks.
   \item Based on HIN Block, we design a multi-stage architecture, HINet, for image restoration tasks, and achieves the state-of-the-art performance with fewer MACs and inference time compares to the SOTA method~\cite{zamir2021multi}.
   \item Extensive experiments are conducted to demonstrate the effectiveness of our proposed HIN Block and HINet. With the help of HIN Block and HINet, we won 1st place on the NTIRE 2021 Image Deblurring Challenge - Track2. JPEG Artifacts~\cite{nah2021ntire}, with a PSNR of 29.70.
\end{itemize}

\section{Related Work}

\subsection{Normalization in low-level computer vision tasks:}
Normalization has become an essential component in high-level computer vision tasks (especially Batch Normalization) but is rarely used in low-level computer vision tasks. ~\cite{nah2017deep} modified the ResBlock~\cite{he2016deep} by removing batch normalization since they trained the model with a mini-batch of size 2 in deblur. ~\cite{lim2017enhanced} removed batch normalization in super-resolution, which the batch normalization get rid of range flexibility from networks.  As the disharmony between image restoration tasks and Batch Normalization (BN) discussed in ~\cite{yu2018wide}, image restoration tasks commonly use small image patches and small mini-batch size to train the network, which causes the statistics of BN unstable. Moreover, the image restoration task is a per-image dense pixel value prediction task, which is scale sensitivity. While BN is usually helpful in scale insensitive tasks. 

In addition to the above, Instance Normalization~\cite{ulyanov2016instance} is proposed to replace Batch Normalization in ~\cite{ulyanov2016texture} to improve the performance of the style transfer task. ~\cite{huang2017arbitrary} demonstrates Instance Normalization is the normalization of low-level features to some extent. They proposed adaptive instance normalization to the style transfer task by aligning the channel-wise statistics in Instance Normalization of style image to content image. Based on ~\cite{huang2017arbitrary}, ~\cite{kim2020transfer} adopts an adaptive instance normalization as a regularizer to build denoiser and transfers knowledge learned from synthetic noise data to the real-noise denoiser.
Unlike ~\cite{kim2020transfer}, we extend the Instance Normalization as a method of feature enhancement and apply it to the image restoration tasks \emph{directly} without transfer learning. 

\subsection{Architectures for Image Restoration }

The single-stage methods are widely used in image restoration tasks, and these methods generally improve the network capacity through the complex network structure~\cite{anwar2020densely, zhang2018density}. The multi-stage methods decompose the complex image restoration task into smaller easier sub-tasks, employing a lightweight subnetwork at each stage. ~\cite{fu2019lightweight} introduce the mature Gaussian-Laplacian image pyramid decomposition technology to the neural network, and uses a relatively shallow network to handle the learning problem at each pyramid level. ~\cite{ren2019progressive} proposes a progressive recurrent network by repeatedly unfolding a shallow ResNet~\cite{he2016deep}, and introduces a recurrent layer to exploit the dependencies of deep features across stages. ~\cite{zhang2019deep} proposes a deep stacked hierarchical multi-patch network. Each level focus on different scales of the blur and the finer level contributes its residual image to the coarser level. ~\cite{zamir2021multi} proposes a multi-stage progressive image restoration architecture, where there are two encoder-decoder subnetworks and one original resolution subnetwork. ~\cite{zamir2021multi} also proposes a supervised attention module (SAM) and a cross-stage feature fusion (CSFF) module between every two stages to enrich the features of the next stage. Our model also uses these two modules to facilitate achieving significant performance gain and uses two simple U-Nets ~\cite{ronneberger2015u} as the subnetworks.

\begin{figure*}
    \centering
    \includegraphics[width=0.96\textwidth]{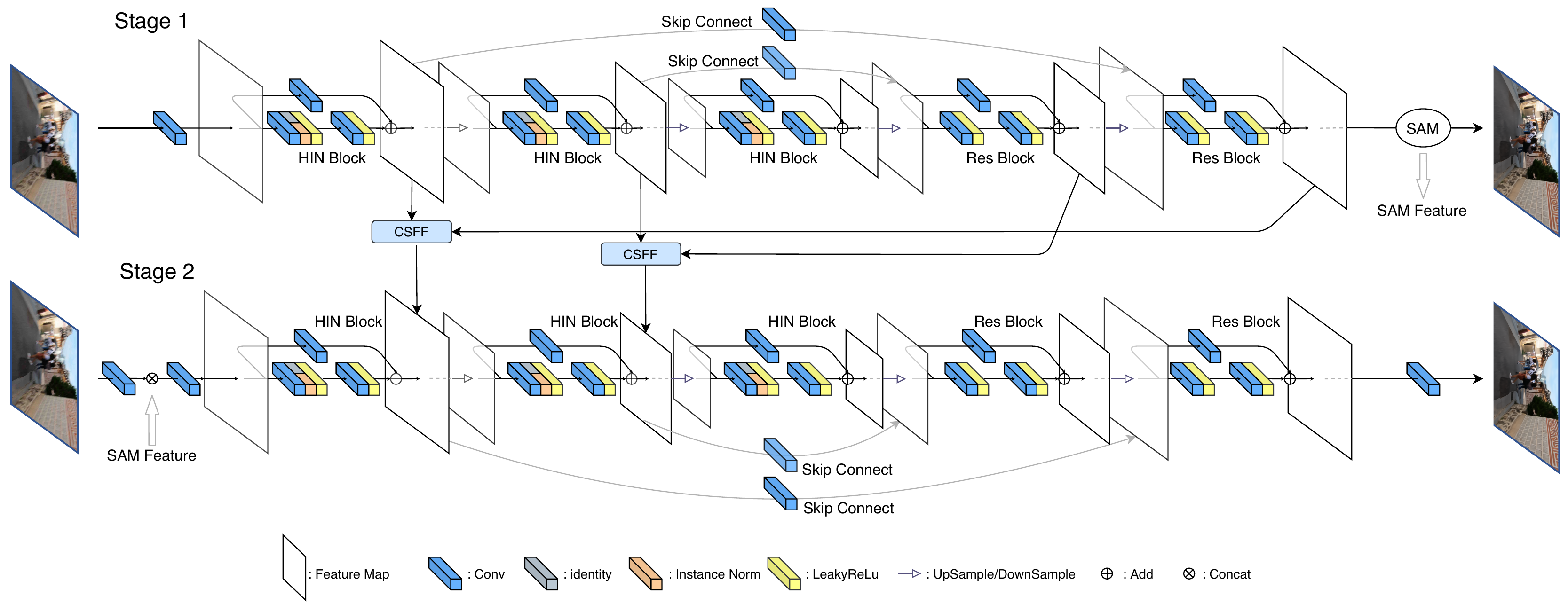}
    \caption{Proposed Half Instance Normalization Network (HINet). The encoder of each subnetwork contains Half Instance Normalization Blocks (HIN Block). For simplicity, we only show 3 layers of HIN Block in the figure, and HINet has a total of 5 layers. We adopt CSFF and SAM modules from MPRNet~\cite{Zamir2021MPRNet}.}
    \label{fig:pipeline}
    \vspace{-0.2cm}
\end{figure*}

\section{Approach}
In this section, we provide more detailed explanations about HINet and HIN Block in the following subsections. Specifically, we introduce HINet in \ref{HINet} and HIN Block in \ref{HIN Block}.

\subsection{HINet}
\label{HINet}
The architecture of our proposed Half Instance Normalization Network (HINet) is shown in Figure~\ref{fig:pipeline}. HINet consists of two subnetworks, each of which is a U-Net ~\cite{ronneberger2015u}. 
As for U-Net in each stage, we use one $3 \times 3$ convolutional layer to extract the initial features. Then those features are input into an encoder-decoder architecture with four downsamplings and upsamplings. We use convolution with kernel size equal to 4 for downsampling, and use transposed convolution with kernel size equal to 2 for upsampling. In the encoder component, we design Half Instance Normalization Blocks to extract features in each scale, and double the channels of features when downsampling. In the decoder component, we use ResBlocks~\cite{he2016deep}  to extract high-level features, and fuse features from the encoder component to compensate for the loss of information caused by resampling. As for ResBlock, we use leaky ReLU~\cite{maas2013rectifier} with a negative slope equal to 0.2 and remove batch normalization. Finally, we get the residual output of the reconstructed image by using one $3 \times 3$ convolution.

We use cross-stage feature fusion (CSFF) module and supervised attention module (SAM) to connect two subnetworks, where these two modules come from~\cite{zamir2021multi}. As for CSFF module, we use $3 \times 3$ convolution to transform the features from one stage to the next stage for aggregation, which helps to enrich the multi-scale features of the next stage. As for SAM, we replace the $1 \times 1$ convolutions in the original module with $3 \times 3$ convolutions and add bias in each convolution. By introducing SAM, the useful features at the current stage can propagate to the next stage and the less informative ones will be suppressed by the attention masks~\cite{zamir2021multi}.

In addition to the network architecture, we use Peak Signal-to-Noise Ratio (PSNR)
as the metric of the loss function, which is PSNR loss. Let $X_i \in \mathbb{R}^{N \times C \times H \times W}$ denotes the input of subnetwork $i$, where $N$ is the batch size of data, $C$ is the number of channels, $H$ and $W$ are spatial size. $R_i \in \mathbb{R}^{N \times C \times H \times W}$ denotes the final predict of subnetwork $i$, and $Y \in \mathbb{R}^{N \times C \times H \times W}$ is the ground truth in each stage. Then we optimize HINet end-to-end as follows:
\begin{equation}
    Loss = -\sum_{i=1}^{2} PSNR((R_i + X_i), Y) \label{eq:dengshi1}
\end{equation}

\begin{figure}
    \centering
    \includegraphics[width=0.47\textwidth]{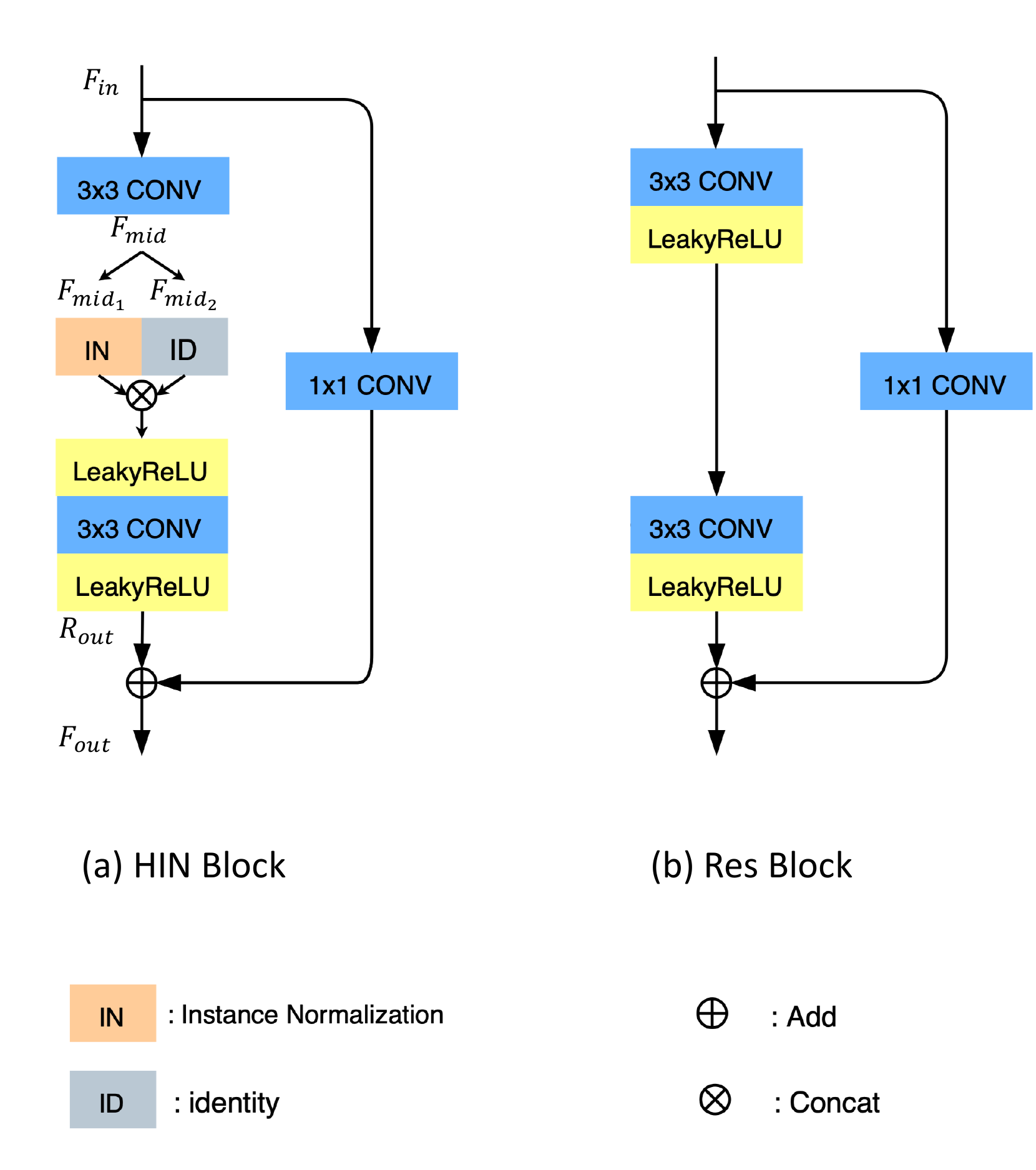}
    \vspace{-0.1cm}
    \caption{Proposed Half Instance Normalization Block (HIN Block) and ResBlock in details.}
    \label{fig:HIN_block}
    \vspace{-0.3cm}
\end{figure}

\subsection{Half Instance Normalization Block}
\label{HIN Block}
Because of variance of small image patches differ a lot among mini-batches and the different formulations of training and testing~\cite{yu2018wide}, BN\cite{ioffe2015batch} is not commonly used in low-level tasks~\cite{ledig2017photo,lim2017enhanced}. Instead, Instance Normalization (IN) keeps the same normalization procedure consistent in both training and inference. Further, IN re-calibrates the mean and variance of features without the influence of batch dimension, which can keep more scale information than BN. We use IN to build Half Instance Normalization Block (HIN block). By introducing HIN block, the modeling capacity of HINet is improved (as shown in Figure~\ref{fig:more_iters}). Moreover, the extra parameters and computational cost introduced by IN can be ignored.

As shown in Figure~\ref{fig:HIN_block} a. HIN block firstly takes the input features $F_{in} \in \mathbb{R}^{C_{in} \times H \times W}$ and generates intermediate features $F_{mid} \in \mathbb{R}^{C_{out} \times H \times W}$ with $3 \times 3$ convolution, where $C_{in}$/$C_{out}$ is the number of input/output channels for HIN block. Then, the features $F_{mid}$ are divided into two parts ($F_{mid_1}/F_{mid_2} \in \mathbb{R}^{C_{out}/2 \times H \times W}$). The first part $F_{mid_1}$ is normalized by IN with learnable affine parameters and then concatenates with $F_{mid_2}$ in channel dimension. HIN blocks use IN on the half of the channels and keep context information by the other half of the channels. Later experiments will also show that this design is more friendly to features in shallow layers of the network. After the concat operation, the residual features $R_{out} \in \mathbb{R}^{C_{out} \times H \times W}$ are obtained by passing features to one $3 \times 3$ convolution layer and two leaky ReLU layers, which is shown in Figure~\ref{fig:HIN_block} a. Finally, HIN blocks output $F_{out}$ by add $R_{out}$ with shortcut features (obtained after $1 \times 1$ convolution).

\section{Experiments}
We evaluate our approach on multiple datasets across image restoration tasks. We report the standard metrics in image restoration including PSNR and SSIM. The datasets used for training are described next.
\subsection{Implementation Details}
\paragraph{Datasets}
As in ~\cite{Zamir2021MPRNet}, we train our models on SIDD~\cite{abdelhamed2018high} for image denoising, GoPro~\cite{nah2017deep} for image deblurring, and 13,712 clean-rain image pairs (for simplicity, denoted as Rain13k in the following) gathered from ~\cite{fu2017removing,li2016rain,yang2017deep,zhang2018density,zhang2019image} for image deraining. In addition, we use REDS~\cite{nah2019ntire} dataset for image deblurring with JPEG artifacts, and we denote it as REDS dataset for simplicity. For evaluation, we follow the setting in the NTIRE 2021 Challenge on Image Deblurring~\cite{nah2021ntire}, \ie use 300 images in the validation set of REDS, denoted as REDS-val-300 next. 

\paragraph{Training}
The networks are trained with Adam optimizer. The learning rate is set to $2\times10^{-4}$ by default, and decreased to $1\times10^{-7}$ with cosine annealing strategy~\cite{loshchilov2016sgdr}. We train our models on $256\times256$ patches with a batch size of 64 for $4\times10^5$ iterations. We apply flip and rotation as data augmentation.
Following ~\cite{zhang2018shufflenet}, we customize the network to the desired complexity by applying a scale factor $s$ on the number of channels, \eg ``HINet $s\times$'' denotes scaling the number of channels in basic HINet $s$ times. 

\begin{table}
\centering
\tablestyle{5pt}{1.05}\setlength{\tabcolsep}{1.mm}\begin{tabular}{l|cc}
& \multicolumn{2}{|c}{SIDD~\cite{abdelhamed2018high}}\\
Method & PSNR & SSIM\\ 
\hline
DnCNN~\cite{zhang2017beyond}&23.66&0.583\\
MLP~\cite{burger2012image}&24.71&0.641\\
DM3D~\cite{dabov2007image}&25.65&0.685\\
CBDNet$^{\color{red}*}$~\cite{guo2019toward}&30.78&0.801\\
RIDNet$^{\color{red}*}$~\cite{anwar2019real}&38.71&0.951\\
AINDNet$^{\color{red}*}$~\cite{kim2020transfer}&38.95&0.952\\
VDN~\cite{yue2019variational}&39.28&0.956\\
SADNet$^{\color{red}*}$~\cite{chang2020spatial}&39.46&0.957\\
DANet+$^{\color{red}*}$~\cite{yue2020dual}&39.47&0.957\\
CycleISP$^{\color{red}*}$~\cite{Zamir2020CycleISP}&39.52&0.957\\
MPRNet~\cite{Zamir2021MPRNet}&39.71&0.958\\
\hline
HINet $0.5\times$(\textbf{ours})&\underline{39.82}&0.958\\
HINet (\textbf{ours})&\textbf{39.99}&0.958\\
\end{tabular}
\vspace{-0.2cm}
\caption{Denoising comparisons on SIDD~\cite{abdelhamed2018high} dataset. $^{\color{red}*}$ denotes the methods that use additional training data. Best and second best scores are \textbf{highlighted} and \underline{underlined}. Our HINet achieves 0.28 dB absolute improvement in PSNR over the previous best method MPRNet~\cite{Zamir2021MPRNet}.}
\label{tab.SIDD.results}
\vspace{-0.2cm}
\end{table}


\begin{table}
\centering
\tablestyle{5pt}{1.05}\setlength{\tabcolsep}{1.mm}\begin{tabular}{l|cc}
& \multicolumn{2}{|c}{GoPro~\cite{nah2017deep}}\\
Method & PSNR & SSIM\\
\hline
Xu et al.~\cite{xu2013unnatural} & 21.00 & 0.741\\
Hyun et al.~\cite{hyun2013dynamic} & 23.64 & 0.824\\
Whyte et al.~\cite{whyte2012non} & 24.60 & 0.846\\
Gong et al.~\cite{gong2017motion} & 26.40 & 0.863\\
DeblurGAN~\cite{kupyn2018deblurgan}& 28.70 & 0.858\\
Nah et al.~\cite{nah2017deep} & 29.08 & 0.914\\
Zhang et al.~\cite{zhang2018dynamic} & 29.19 & 0.931\\
DeblurGAN-v2~\cite{kupyn2019deblurgan} & 29.55 & 0.934\\
SRN~\cite{tao2018scale} & 30.26 & 0.934\\
Gao et al.~\cite{gao2019dynamic} & 30.90 & 0.935\\
DBGAN~\cite{zhang2020deblurring} & 31.10 & 0.942\\
MT-RNN~\cite{park2020multi} & 31.15 & 0.945\\
DMPHN~\cite{Zhang_2019_CVPR} & 31.20 & 0.940\\
Suin et al.~\cite{suin2020spatially} & 31.85 & 0.948\\
MPRNet~\cite{Zamir2021MPRNet} & \underline{32.66} & \textbf{0.959}\\
\hline
HINet (\textbf{ours}) & \textbf{32.77} & \textbf{0.959}\\
\end{tabular}
\vspace{-0.2cm}
\caption{Deblurring comparisons on GoPro~\cite{nah2017deep} dataset. Best and second best scores are \textbf{highlighted} and \underline{underlined}. Our HINet achieves 0.11 dB absolute improvement in PSNR over the previous best method MPRNet~\cite{Zamir2021MPRNet}.}
\label{tab.GoPro.results}
\vspace{-0.4cm}
\end{table}

\begin{table*}
\centering
\tablestyle{5pt}{1.05}\setlength{\tabcolsep}{1.mm}\begin{tabular}{l|cccccccccc|cc}
& \multicolumn{2}{c}{Test100~\cite{zhang2019image}}& \multicolumn{2}{c}{Rain100H~\cite{yang2017deep}}& \multicolumn{2}{c}{Rain100L~\cite{yang2017deep}}& \multicolumn{2}{c}{Test2800~\cite{fu2017removing}}& \multicolumn{2}{c}{Test1200~\cite{zhang2018density}}& \multicolumn{2}{|c}{Average}\\
Method&PSNR & SSIM&PSNR &SSIM &PSNR & SSIM & PSNR & SSIM & PSNR & SSIM & PSNR & SSIM\\
\hline
DerainNet~\cite{fu2017clearing}&22.77 & 0.810 & 14.92 & 0.592 & 27.03 & 0.884 & 24.31 & 0.861 & 23.38 & 0.835 & 22.48 & 0.796\\
SEMI~\cite{wei2019semi} & 22.35 & 0.788 & 16.56 & 0.486 & 25.03 & 0.842 & 24.43 & 0.782 & 26.05 & 0.822 & 22.88 & 0.744\\
DIDMDN~\cite{zhang2018density} & 22.56 & 0.818 & 17.35 & 0.524 & 25.23 & 0.741 & 28.13 & 0.867 & 29.65 & 0.901 & 24.58 & 0.770\\
UMRL~\cite{yasarla2019uncertainty} & 24.41 & 0.829 & 26.01 & 0.832 & 29.18 & 0.923 & 29.97 & 0.905 & 30.55 & 0.910 & 28.02 & 0.880\\
RESCAN~\cite{li2018recurrent}& 25.00 & 0.835 & 26.36 & 0.786 & 29.80 & 0.881 & 31.29 & 0.904 & 30.51 & 0.882 & 28.59 & 0.857\\
PreNe~\cite{ren2019progressive} & 24.81 & 0.851 & 26.77 & 0.858 & 32.44 & 0.950 & 31.75 & 0.916 & 31.36 & 0.911 & 29.42 & 0.897\\
MSPFN~\cite{jiang2020multi} & 27.50 & 0.876 & 28.66 & 0.860 & 32.40 & 0.933 & 32.82 & 0.930 & 32.39 & 0.916 & 30.75 & 0.903\\
MPRNet~\cite{Zamir2021MPRNet} & \underline{30.27} & \underline{0.897} & \underline{30.41} & \underline{0.890} & \underline{36.40} & \underline{0.965} & \underline{33.64} & \underline{0.938} & \underline{32.91} & \underline{0.916} & \underline{32.73} & \underline{0.921}\\
\hline
HINet (\textbf{ours})& \textbf{30.29}& \textbf{0.906} & \textbf{30.65} & \textbf{0.894} & \textbf{37.28} & \textbf{0.970} & \textbf{33.91} & \textbf{0.941} & \textbf{33.05} & \textbf{0.919} & \textbf{33.03} & \textbf{0.926}\\
\end{tabular}
\vspace{-.2cm}
\caption{Deraining comparisons on Test100~\cite{zhang2019image}, Rain100H~\cite{yang2017deep}, Rain100L~\cite{yang2017deep}, Test2800~\cite{fu2017removing} and Test1200~\cite{zhang2018density}. In addition, the average results over these datasets are provided. Best and second best scores are \textbf{highlighted} and \underline{underlined}. Our HINet achieves 0.3 dB absolute improvement in PSNR over the previous best method MPRNet~\cite{Zamir2021MPRNet}.}
\label{tab.Rain13k.results}
\vspace{-0.3cm}
\end{table*}



\begin{table}
\centering
\tablestyle{5pt}{1.05}\setlength{\tabcolsep}{1.mm}\begin{tabular}{l|l|c|rr|rr}
Dataset& Method & PSNR & \multicolumn{2}{c}{MACs(G)} &Time(ms)&speedup\\
\hline
\multirow{3}{4em}{SIDD~\cite{abdelhamed2018high}} & MPRNet~\cite{Zamir2021MPRNet} & 39.71 & 573.50&100\% & 78.8 & $1\times$\\
&HINet $0.5\times$&39.82&42.88&7.5\% & 11.6 & $6.8\times$\\
&HINet &39.99&170.71&29.8\% & 27.0& $2.9\times$\\
\hline
\multirow{2}{4em}{REDS~\cite{nah2019ntire}} & MPRNet~\cite{Zamir2021MPRNet} & 28.81 & 760.11&100\% & 90.1& $1\times$\\
&HINet  &28.83&170.71&22.5\% & 27.0 & $3.3\times$\\
\hline
\multirow{2}{4em}{GoPro~\cite{nah2017deep}} & MPRNet~\cite{Zamir2021MPRNet} & 32.66 & 760.11&100\% & 90.1 & $1\times$ \\
&HINet  &32.77&170.71&22.5\% & 27.0 & $3.3\times$\\
\hline
\multirow{2}{4em}{Rain13k} & MPRNet~\cite{Zamir2021MPRNet} & 32.73 &141.28&100\% & 37.4 & $1\times$\\
&HINet  &33.03&170.71&120.8\% & 27.0 & $1.4\times$\\
\end{tabular}
\vspace{-.2cm}
\caption{Comparing the PSNR and MACs of MPRNet~\cite{Zamir2021MPRNet} and ours. For Rain13k, we compare the average PSNR over Test100~\cite{zhang2019image}, Rain100H~\cite{yang2017deep}, Rain100L~\cite{yang2017deep}, Test2800~\cite{fu2017removing} and Test1200~\cite{zhang2018density}. MACs and Time are estimated with the input size of $1\times3\times256\times256$. The proportion of the calculations and the speedup compared to MPRNet~\cite{Zamir2021MPRNet} are also listed. Runtimes are computed with the Tesla V100 GPU.}
\label{tab.GmacsCompare}
\vspace{-.3cm}
\end{table}

\subsection{Main Results}
\begin{figure*}[t]
    \centering
    \includegraphics[width=0.87\textwidth]{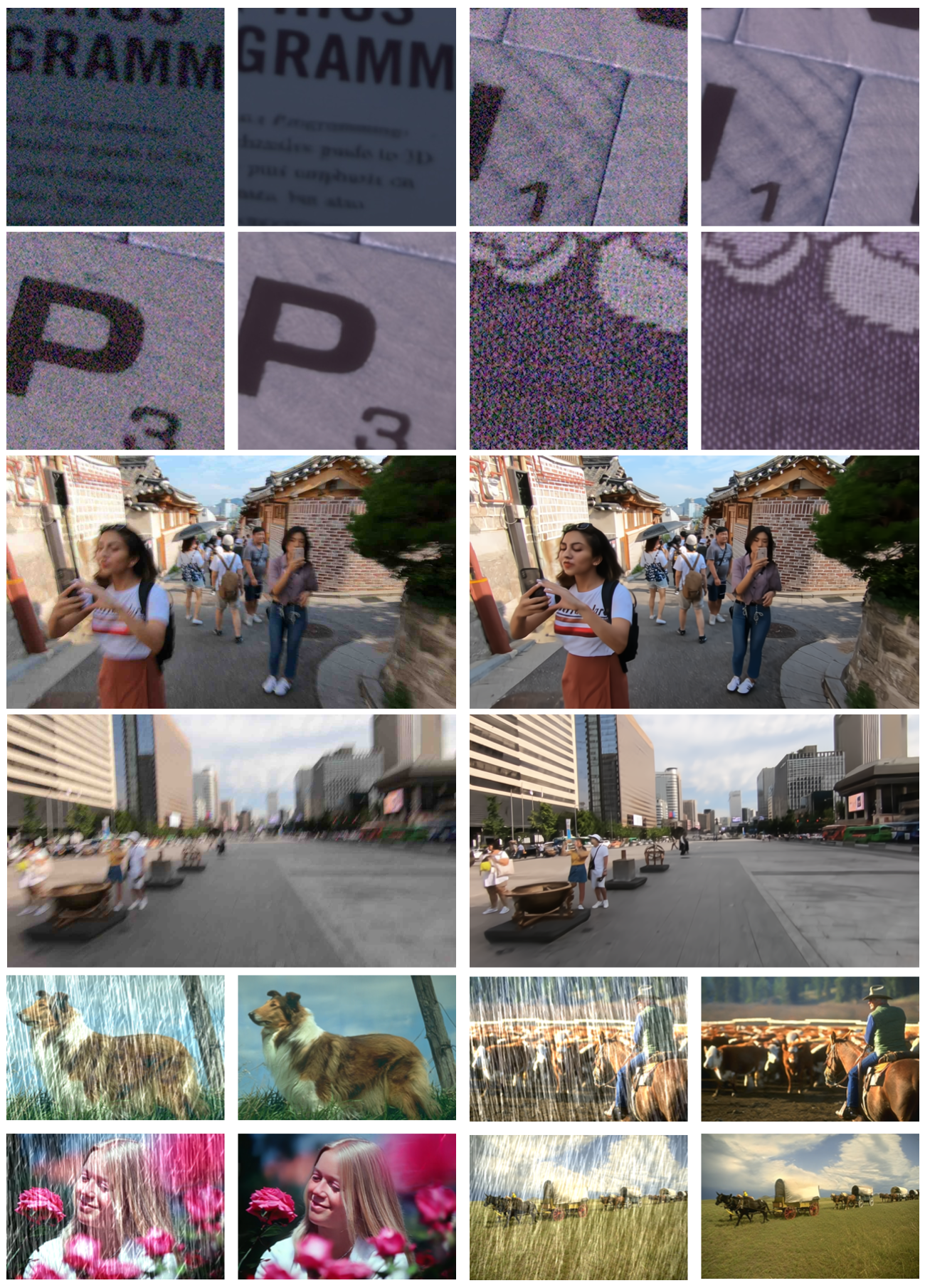}
    \caption{More visualized results of HINet on various image restoration tasks. For each image pair, the left one is degraded and the right one is predicted by HINet.}
    \label{fig:visual_more}
\end{figure*}

We show the effectiveness of HINet on different datasets in Table~\ref{tab.SIDD.results}, Table~\ref{tab.GoPro.results} and Table~\ref{tab.Rain13k.results}. In addition, we compare the MACs (\ie multiplier-accumulator operations) and inference time of MPRNet~\cite{Zamir2021MPRNet} and HINet in Table~\ref{tab.GmacsCompare}. MACs is estimated when the input is $1\times3\times256\times256$. Moreover, we conduct quality experiments to show the superiority of our method as shown in Figure~\ref{fig:visual_more}.

\paragraph{SIDD~\cite{abdelhamed2018high}} For image denoising, we train our model on the 320 high-resolution images of the SIDD dataset, and test on the 1280 patches from SIDD dataset. The results are shown in Table~\ref{tab.SIDD.results} and Table~\ref{tab.GmacsCompare}. Surprisingly, using only 7.5\% of MPRNet~\cite{Zamir2021MPRNet}'s MACs, our model exceeds it by 0.11 dB in PSNR. Moreover, our model exceeds MPRNet a big margin, 0.28 dB in PSNR under 30\% of the MACs, and is 2.9 times faster than MPRNet.

\paragraph{REDS~\cite{nah2019ntire} and GoPro~\cite{nah2017deep}} For image deblurring, we train our model on REDS~\cite{nah2019ntire} dataset with jpeg compression artifacts, and evaluate the results on REDS-val-300 as we described above. In addition, we conduct experiments on GoPro~\cite{nah2017deep} dataset for image deblurring following ~\cite{Zamir2021MPRNet,Zhang_2019_CVPR} \etc. It contains 2103 image pairs for training and 1111 pairs for evaluation. In Table~\ref{tab.GoPro.results} and Table~\ref{tab.GmacsCompare}, we compare our approach to the state-of-the-art methods. We get a comparable performance to MPRNet~\cite{Zamir2021MPRNet} with only 22.5\% MACs and $3.3\times$ speed advantage. It indicates the efficiency of our model. 

\paragraph{Rain13k} For image deraining, we train our model on Rain13k as described above and evaluate results by Y channel in YCbCr color space following ~\cite{jiang2020multi,Zamir2021MPRNet}. It contains 13712 image pairs in the training set, and we evaluate the results on Test100~\cite{zhang2019image}, Rain100H~\cite{yang2017deep}, Rain100L~\cite{yang2017deep}, Test2800~\cite{fu2017removing}, Test1200~\cite{zhang2018density}. We show the results in Table~\ref{tab.Rain13k.results}. HINet achieves 0.3 dB absolute improvement in PSNR over the previous best method MPRNet~\cite{Zamir2021MPRNet} and $1.4\times$ faster than it.

\subsection{Ablation}The core idea of HINet lies in HIN Block. We evaluate it from multiple perspectives in this subsection.
It should be noted that these experiments are \emph{not} to achieve the performance of the state-of-the-art, but to illustrate the superiority of HIN on various models and image restoration tasks.
Therefore we mainly use the UNet of the first stage of HINet $0.5\times$, without skip connections between encoder and decoder for ablation experiments for fast feedback, and we denoted the UNet as ``HINet Simple'' in next. It is trained on $512\times512$ patches with a batch size of $32$ for $3\times10^5$ iterations. For optimizer, we follow the settings in ~\cite{Zamir2021MPRNet} except we set the learning rate to $4\times10^{-4}$ instead of $2\times10^{-4}$.

\paragraph{The effectiveness of Half Instance Normalization:} 
We conduct experiments on various models and datasets, as we shown in Table~\ref{hin.effectiveness}. 

On REDS~\cite{nah2019ntire} dataset (shown in Table~\ref{hin.effectiveness.REDS}), HIN brings 0.12 dB in PSNR for HINet Simple. 
This illustrates the effectiveness of HIN on the REDS~\cite{nah2019ntire} dataset and HINet Simple model.

On GoPro~\cite{nah2017deep} dataset (shown in Table~\ref{hin.effectiveness.GoPro}), we reimplement the DMPHN(1-2-4-8)~\cite{Zhang_2019_CVPR}. We set the learning rate to $2\times10^{-4}$ and decreased to $1\times10^{-6}$ with cosine annealing strategy~\cite{loshchilov2016sgdr}. We train the model on $256\times256$ patches, with a batch size of 32 for $2\times10^5$ iterations. Flip and rotation are applied as data augmentation. HIN brings 0.42 dB boost in PSNR, it demonstrates HIN is effective on different models and different datasets.

We train PRMID~\cite{wang2020practical} and CycleISP~\cite{Zamir2020CycleISP} on SIDD~\cite{abdelhamed2018high} dataset as shown in Table~\ref{hin.effectiveness.SIDD}. For PRMID, we set the initial learning rate to $3\times10^{-3}$ with a batch size of 16 for $4\times10^5$ iterations. For CycleISP, we set the initial learning rate to $2\times10^{-4}$ with a batch size of 16 for $3\times10^5$ iterations. Adam optimizer, cosine annealing strategy~\cite{loshchilov2016sgdr} and flip/rotation as data augmentation are adopted in both cases. HIN brings 0.09 dB and 0.06 dB boost in PSNR on PRMID and CycleISP respectively. Since SIDD~\cite{abdelhamed2018high} and GoPro~\cite{nah2017deep}/REDS~\cite{nah2019ntire} are different image restoration task datasets, it demonstrates that HIN is effective in different image restoration tasks.

\begin{table}
    \tablestyle{5pt}{1.05}\setlength{\tabcolsep}{1.mm}\begin{subtable}[h]{0.49\textwidth}
    \centering
      \begin{tabular}{lc|cc}
      &&\multicolumn{2}{c}{REDS~\cite{nah2019ntire}}\\
      Method&HIN?&PSNR&SSIM\\
      \hline
      HINet Simple&- &28.11&0.847\\
      HINet Simple&\checkmark &28.23&0.850\\
      \end{tabular}
      \vspace{.1cm}
      \caption{Comparison of the models with/without HIN on REDS~\cite{nah2019ntire} dataset for deblurring. HIN brings 0.12 dB in PSNR to HINet Simple. }
      \label{hin.effectiveness.REDS}
   \end{subtable}
   \begin{subtable}[h]{0.49\textwidth}
   \vspace{.2cm}
   \centering
   \tablestyle{5pt}{1.05}\setlength{\tabcolsep}{1.mm}\begin{tabular}{lc|cc}
      &&\multicolumn{2}{c}{GoPro~\cite{nah2017deep}}\\
      Method&HIN?&PSNR&SSIM\\
      \hline
      DMPHN(1-2-4-8)~\cite{Zhang_2019_CVPR}&- &30.98&0.943\\
      DMPHN(1-2-4-8)~\cite{Zhang_2019_CVPR}&\checkmark &31.40&0.948\\
      \end{tabular}
      \vspace{.1cm}
      \caption{Comparison of the models with/without HIN on GoPro~\cite{nah2017deep} dataset for deblurring. HIN brings 0.42 dB in PSNR to DMPHN(1-2-4-8)~\cite{Zhang_2019_CVPR}. It indicates HIN's effectiveness is robust to datasets and models.}
      \label{hin.effectiveness.GoPro}
   \end{subtable}
   \vspace{.15cm}
   \begin{subtable}[h]{0.49\textwidth}
   \centering
   \vspace{.25cm}
   \tablestyle{5pt}{1.05}\setlength{\tabcolsep}{1.mm}\begin{tabular}{lc|c}
   \vspace{.1cm}
   &&\multicolumn{1}{c}{SIDD~\cite{abdelhamed2018high}}\\
      Method&HIN?&PSNR\\
      \hline
      PRMID~\cite{wang2020practical}&- &39.30\\
      PRMID~\cite{wang2020practical}&\checkmark&39.39\\
      \hline
      CycleISP~\cite{Zamir2020CycleISP}&- &39.50\\
      CycleISP~\cite{Zamir2020CycleISP}&\checkmark&39.56\\
   \end{tabular}
   \vspace{.1cm}
   \caption{Comparison of the models with/without HIN on SIDD~\cite{abdelhamed2018high} dataset for denoising. HIN brings 0.09 dB and 0.06 dB in PSNR to PRMID~\cite{wang2020practical} and CycleISP~\cite{Zamir2020CycleISP} respectively. It indicates HIN's effectiveness is robust to image restoration tasks and model size.}
   \label{hin.effectiveness.SIDD}
   \end{subtable}
\vspace{-.1cm}
\caption{To demonstrate the effectiveness of HIN, we conduct experiments on various datasets and models.}
\vspace{-.2cm}
\label{hin.effectiveness}
\end{table}

\paragraph{Comparison with other Normalizations:}
Normalization has not been fully explored in image restoration tasks. We compare HIN with other normalizations to demonstrate the superiority of our proposed HIN. We conduct experiments on REDS~\cite{nah2019ntire} dataset on HINet Simple. The results are shown in Table~\ref{tab:vs.norm}. We denote the HINet Simple w/o. Norm as the baseline in the following. BN~\cite{ioffe2015batch} results a significant performance drop compares to baseline, \ie 0.12 dB in PSNR. We conjure that is because of the inaccurate batch statistics estimation when batch size is small. It is alleviated by SyncBN~\cite{zhang2018context} in some extent, \ie HINet Simple w/. SyncBN brings 0.1 dB gain compares to HINet Simple w/. BN. However, it is still inferior to baseline (28.09 dB vs. 28.11 dB in PSNR). As we can see, with the help of IN~\cite{ulyanov2017improved}, HINet Simple w/. IN exceed baseline by 0.03 dB in PSNR. It indicates IN facilitates the training of image restoration tasks. As shown in Table~\ref{tab:vs.norm}, HIN exceeds its counterparts, and it brings 0.12 dB gain in PSNR compares to baseline. It demonstrates the superiority of our proposed HIN.

\begin{table}[h]
    \centering
    \tablestyle{5pt}{1.05}\setlength{\tabcolsep}{1.mm}\begin{tabular}{l|c}
        Method &PSNR \\
        \hline
        HINet Simple w/o. Norm & 28.11 \\
        HINet Simple w/. BN~\cite{ioffe2015batch} & 27.99\\
        HINet Simple w/. SyncBN~\cite{zhang2018context} &28.09\\
        HINet Simple w/. LN~\cite{ba2016layer} & 28.09 \\
        HINet Simple w/. IN~\cite{ulyanov2017improved} & 28.14 \\
        HINet Simple w/. IBN~\cite{pan2018two} & 28.18 \\
        HINet Simple w/. GN~\cite{wu2018group} & 28.19\\
        \hline
        HINet Simple w/. HIN(\textbf{ours}) & \textbf{28.23}\\
    \end{tabular}
    \caption{Comparison of different normalization approaches on the image restoration task. Experiments are conducted on REDS~\cite{nah2019ntire} dataset. HINet Simple w/. BN means a fully normalized model with BN in the encoder.}
    \label{tab:vs.norm}
\end{table}

\begin{figure}
    \centering
    \includegraphics[width=0.48\textwidth]{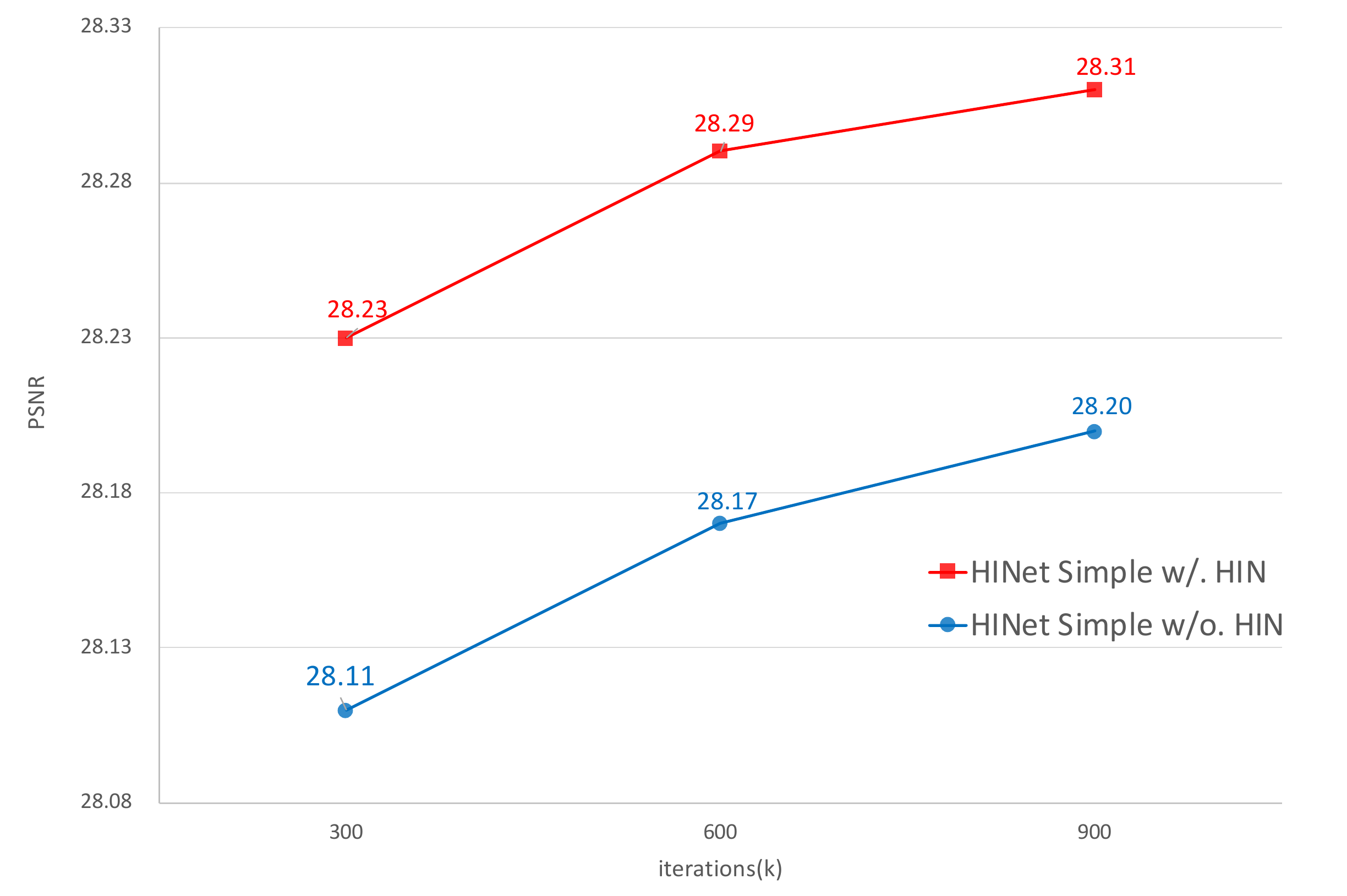}
    \caption{Effectiveness of HIN when training more iterations.}
    \label{fig:more_iters}
\end{figure}

\paragraph{More training iterations:}
We analyze the impact of increasing the number of training iterations based on HINet Simple. We train the model on REDS~\cite{nah2019ntire} dataset for 300k, 600k, and 900k iterations with/without HIN respectively. The results are shown in Figure~\ref{fig:more_iters}. The gap between HINet Simple w/. HIN and HINet Simple w/o. HIN does not decrease as the number of iterations increased. We conjure that this is because HIN increases the upper limit of the model, not just speeds up the convergence.

\paragraph{Guideline of add HIN layer in an existing network:}
HINet Simple consists of 5 encoder blocks and 5 decoder blocks. We further explore the appropriate add location of HIN Block. The results are shown in Table~\ref{tab:HIN.position}. It indicates that adding HIN to all encoder blocks gets the highest score. In addition, adding HIN to the encoder and decoder causes performance drop \ie 28.23 dB to 28.21 dB. It demonstrates that adding more HIN does not necessarily lead to better performance. In practice, add one HIN layer to each encoder block might be a good choice. 

\begin{table}[h]
    \centering
    \tablestyle{5pt}{1.05}\setlength{\tabcolsep}{1.mm}\begin{tabular}{l|ccccc|c|c}
    &\multicolumn{5}{c|}{Encoder}&\\
    Method&1&2&3&4&5&Decoder&PSNR\\
         \hline
         \multirow{12}{6em}{HINet Simple} &-&-&-&-&-&-&28.11\\
         &\checkmark&\checkmark&-&-&-&-&28.15\\
         &-&\checkmark&\checkmark&-&-&-&28.19\\
         &-&-&\checkmark&\checkmark&-&-&28.19\\
         &-&-&-&\checkmark&\checkmark&-&28.21\\
         &\checkmark&\checkmark&\checkmark&-&-&-&28.19\\
         &-&\checkmark&\checkmark&\checkmark&-&-&28.20\\
         &-&-&\checkmark&\checkmark&\checkmark&-&28.21\\
         &\checkmark&\checkmark&\checkmark&\checkmark&-&-&28.21\\
         &-&\checkmark&\checkmark&\checkmark&\checkmark&-&28.22\\
         &\checkmark&\checkmark&\checkmark&\checkmark&\checkmark&-&\textbf{28.23}\\
         &\checkmark&\checkmark&\checkmark&\checkmark&\checkmark&\checkmark&28.21\\
    \end{tabular}
    \caption{Guideline of add HIN layer in an existing network (\eg HINet Simple): adding HIN to all encoder blocks gets highest score, while more HIN does not necessarily lead to better performance.}
    \label{tab:HIN.position}
\end{table}

\subsection{Extension to HINet:}
In order to achieve better performance on NTIRE 2021 Image Deblurring Challenge Track2. JPEG artifacts~\cite{nah2021ntire}, we extend HINet, and adopt test time augmentation strategy. To further enhance the performance, we ensemble 3 similar models. In this subsection, we discuss the impact of these three methods on the results. And at the end, the results of the development phase and the test phase are provided. The results are evaluated on REDS-val-300, except the test phase result.

\paragraph{Wider, Deeper:} 
It has been demonstrated that scaling up the model from width and depth improves the model capacity~\cite{tan2019efficientnet,yu2018wide}. For width, we simply use HINet $2\times$. For depth, we add two residual blocks at the end of each encoder block and decoder block.
It achieves a PSNR of 29.05 dB on REDS-val-300.

\paragraph{Test Time Augmentation and Ensemble:}
We adopt flip and rotation as test time augmentation. It brings about 0.14 dB in PSNR. In addition, we randomly crop hundreds of patches, randomly adopt flip and rotation augmentation on them. It brings about 0.05 dB in PSNR. We simply average the predictions of 3 models as a model ensemble. It brings about 0.01 dB in PSNR. With these strategies, our model boost PSNR from 29.05 dB to 29.25 dB.

\paragraph{Development phase result:}
For the development phase, we randomly crop 720 patches. The results are shown in Table~\ref{result.devphase}.

\begin{table}[]
    \centering
    \tablestyle{5pt}{1.05}\setlength{\tabcolsep}{1.mm}\begin{tabular}{l|ccc}
    Participants & PSNR & SSIM & rank\\
    \hline
        \textbf{ours} & \textbf{29.25} & \textbf{0.8190} & 1 \\
         participant A & \underline{29.17} & \underline{0.8183} & 2\\
         participant B & 29.14 & 0.8170 & 3\\
         participant C & 29.11 & 0.8171 & 4\\
         participant D & 29.10 & 0.8165 & 5\\
         participant E & 29.01 & 0.8141 & 6\\
         participant F & 28.94 & 0.8145 & 7\\
         participant G & 28.75 & 0.8093 & 8\\
         participant H & 28.68 & 0.8103 & 9\\
         participant I & 28.66 & 0.8082 & 10\\
    \end{tabular}
    \vspace{-0.1cm}
    \caption{Development phase result of NTIRE 2021 Image Deblurring Challenge Track 2. JPEG artifacts~\cite{nah2021ntire}. Best and second best scores are \textbf{highlighted} and \underline{underlined}. Our proposed method outperform others by 0.08 dB in PSNR.}
    \label{result.devphase}
\end{table}

\paragraph{Test phase result:}
For the test phase, we randomly crop 1000 patches. The results are shown in ~\ref{result.testphase}.

\begin{table}\centering
\tablestyle{5pt}{1.05}\setlength{\tabcolsep}{1.mm}\begin{tabular}{l|ccc}
Participants & PSNR & SSIM & rank\\
\hline
\textbf{ours} & \textbf{29.70} & \underline{0.8403} & 1\\
Noah\_CVLab & \underline{29.62} & 0.8397 & 2\\
CAPP\_OB & 29.60 & 0.8398 & 3\\
Baidu & 29.59 & 0.8381 & 4\\
SRC-B & 29.56 & 0.8385 & 5\\
Mier & 29.34 & 0.8355 & 6\\
VIDAR & 29.33 & \textbf{0.8565} & 7\\
DuLang & 29.17 & 0.8325 & 8\\
TeamInception & 29.11 & 0.8292 & 9\\
Giantpandacv & 29.07 & 0.8286 & 10\\
Maradona & 28.96 & 0.8264 & 11\\
LAB FHD & 28.92 & 0.8259 & 12\\
SYJ & 28.81 & 0.8222 & 13\\
Dseny & 28.26 & 0.8081 & 14\\
IPCV IITM & 27.91 & 0.8028 & 15\\
DMLAB & 27.84 & 0.8013 & 16\\
Blur Attack & 27.41 & 0.7887 & 17\\
\end{tabular}
\vspace{-0.1cm}
\caption{NTIRE 2021 Image Deblurring Challenge Track 2. JPEG artifacts result~\cite{nah2021ntire}. Best and second best scores are \textbf{highlighted} and \underline{underlined}. We exceed other participants over 0.08 dB in PSNR.}
\label{result.testphase}
\end{table}

\section{Conclusion}

In this work, we reuse Normalization in image restoration tasks. Specifically, we introduce Instance Normalization into a residual block and design an effective and efficient block: Half Instance Normalization Block (HIN Block). In HIN Block, we apply Instance Normalization for half of the intermediate features and keep the content information at the same time. Based on HIN Block, we further propose a multi-stage network called HINet. Between each stage, we use feature fusion and attention-guided map~\cite{zamir2021multi} across stages to ease the flow of information and enhance the multi-scale feature expression. Our proposed HINet surpasses the SOTA on various image restoration tasks. In addition, by using HINet, we won 1st place on the NTIRE 2021 Image De-blurring Challenge - Track2. JPEG Artifacts~\cite{nah2021ntire}.

\clearpage
{\small
\bibliographystyle{ieee_fullname}
\bibliography{egbib}

\begin{thebibliography}{10}\itemsep=-1pt

\bibitem{abdelhamed2018high}
Abdelrahman Abdelhamed, Stephen Lin, and Michael~S Brown.
\newblock A high-quality denoising dataset for smartphone cameras.
\newblock In {\em Proceedings of the IEEE Conference on Computer Vision and
  Pattern Recognition}, pages 1692--1700, 2018.

\bibitem{anwar2019real}
Saeed Anwar and Nick Barnes.
\newblock Real image denoising with feature attention.
\newblock In {\em Proceedings of the IEEE/CVF International Conference on
  Computer Vision}, pages 3155--3164, 2019.

\bibitem{anwar2020densely}
Saeed Anwar and Nick Barnes.
\newblock Densely residual laplacian super-resolution.
\newblock {\em IEEE Transactions on Pattern Analysis and Machine Intelligence},
  2020.

\bibitem{ba2016layer}
Jimmy~Lei Ba, Jamie~Ryan Kiros, and Geoffrey~E Hinton.
\newblock Layer normalization.
\newblock {\em arXiv preprint arXiv:1607.06450}, 2016.

\bibitem{burger2012image}
Harold~C Burger, Christian~J Schuler, and Stefan Harmeling.
\newblock Image denoising: Can plain neural networks compete with bm3d?
\newblock In {\em 2012 IEEE conference on computer vision and pattern
  recognition}, pages 2392--2399. IEEE, 2012.

\bibitem{carion2020end}
Nicolas Carion, Francisco Massa, Gabriel Synnaeve, Nicolas Usunier, Alexander
  Kirillov, and Sergey Zagoruyko.
\newblock End-to-end object detection with transformers.
\newblock In {\em European Conference on Computer Vision}, pages 213--229.
  Springer, 2020.

\bibitem{chang2020spatial}
Meng Chang, Qi Li, Huajun Feng, and Zhihai Xu.
\newblock Spatial-adaptive network for single image denoising.
\newblock In {\em European Conference on Computer Vision}, pages 171--187.
  Springer, 2020.

\bibitem{dabov2007image}
Kostadin Dabov, Alessandro Foi, Vladimir Katkovnik, and Karen Egiazarian.
\newblock Image denoising by sparse 3-d transform-domain collaborative
  filtering.
\newblock {\em IEEE Transactions on image processing}, 16(8):2080--2095, 2007.

\bibitem{fu2017clearing}
Xueyang Fu, Jiabin Huang, Xinghao Ding, Yinghao Liao, and John Paisley.
\newblock Clearing the skies: A deep network architecture for single-image rain
  removal.
\newblock {\em IEEE Transactions on Image Processing}, 26(6):2944--2956, 2017.

\bibitem{fu2017removing}
Xueyang Fu, Jiabin Huang, Delu Zeng, Yue Huang, Xinghao Ding, and John Paisley.
\newblock Removing rain from single images via a deep detail network.
\newblock In {\em Proceedings of the IEEE Conference on Computer Vision and
  Pattern Recognition}, pages 3855--3863, 2017.

\bibitem{fu2019lightweight}
Xueyang Fu, Borong Liang, Yue Huang, Xinghao Ding, and John Paisley.
\newblock Lightweight pyramid networks for image deraining.
\newblock {\em IEEE transactions on neural networks and learning systems},
  31(6):1794--1807, 2019.

\bibitem{gao2019dynamic}
Hongyun Gao, Xin Tao, Xiaoyong Shen, and Jiaya Jia.
\newblock Dynamic scene deblurring with parameter selective sharing and nested
  skip connections.
\newblock In {\em Proceedings of the IEEE/CVF Conference on Computer Vision and
  Pattern Recognition}, pages 3848--3856, 2019.

\bibitem{gong2017motion}
Dong Gong, Jie Yang, Lingqiao Liu, Yanning Zhang, Ian Reid, Chunhua Shen, Anton
  Van Den~Hengel, and Qinfeng Shi.
\newblock From motion blur to motion flow: A deep learning solution for
  removing heterogeneous motion blur.
\newblock In {\em Proceedings of the IEEE conference on computer vision and
  pattern recognition}, pages 2319--2328, 2017.

\bibitem{guo2019toward}
Shi Guo, Zifei Yan, Kai Zhang, Wangmeng Zuo, and Lei Zhang.
\newblock Toward convolutional blind denoising of real photographs.
\newblock In {\em Proceedings of the IEEE/CVF Conference on Computer Vision and
  Pattern Recognition}, pages 1712--1722, 2019.

\bibitem{he2016deep}
Kaiming He, Xiangyu Zhang, Shaoqing Ren, and Jian Sun.
\newblock Deep residual learning for image recognition.
\newblock In {\em Proceedings of the IEEE conference on computer vision and
  pattern recognition}, pages 770--778, 2016.

\bibitem{huang2017arbitrary}
Xun Huang and Serge Belongie.
\newblock Arbitrary style transfer in real-time with adaptive instance
  normalization.
\newblock In {\em Proceedings of the IEEE International Conference on Computer
  Vision}, pages 1501--1510, 2017.

\bibitem{hyun2013dynamic}
Tae Hyun~Kim, Byeongjoo Ahn, and Kyoung Mu~Lee.
\newblock Dynamic scene deblurring.
\newblock In {\em Proceedings of the IEEE International Conference on Computer
  Vision}, pages 3160--3167, 2013.

\bibitem{ioffe2015batch}
Sergey Ioffe and Christian Szegedy.
\newblock Batch normalization: Accelerating deep network training by reducing
  internal covariate shift.
\newblock In {\em International conference on machine learning}, pages
  448--456. PMLR, 2015.

\bibitem{jiang2020multi}
Kui Jiang, Zhongyuan Wang, Peng Yi, Chen Chen, Baojin Huang, Yimin Luo, Jiayi
  Ma, and Junjun Jiang.
\newblock Multi-scale progressive fusion network for single image deraining.
\newblock In {\em Proceedings of the IEEE/CVF Conference on Computer Vision and
  Pattern Recognition}, pages 8346--8355, 2020.

\bibitem{kim2020transfer}
Yoonsik Kim, Jae~Woong Soh, Gu~Yong Park, and Nam~Ik Cho.
\newblock Transfer learning from synthetic to real-noise denoising with
  adaptive instance normalization.
\newblock In {\em Proceedings of the IEEE/CVF Conference on Computer Vision and
  Pattern Recognition}, pages 3482--3492, 2020.

\bibitem{kupyn2018deblurgan}
Orest Kupyn, Volodymyr Budzan, Mykola Mykhailych, Dmytro Mishkin, and
  Ji{\v{r}}{\'\i} Matas.
\newblock Deblurgan: Blind motion deblurring using conditional adversarial
  networks.
\newblock In {\em Proceedings of the IEEE conference on computer vision and
  pattern recognition}, pages 8183--8192, 2018.

\bibitem{kupyn2019deblurgan}
Orest Kupyn, Tetiana Martyniuk, Junru Wu, and Zhangyang Wang.
\newblock Deblurgan-v2: Deblurring (orders-of-magnitude) faster and better.
\newblock In {\em Proceedings of the IEEE/CVF International Conference on
  Computer Vision}, pages 8878--8887, 2019.

\bibitem{ledig2017photo}
Christian Ledig, Lucas Theis, Ferenc Husz{\'a}r, Jose Caballero, Andrew
  Cunningham, Alejandro Acosta, Andrew Aitken, Alykhan Tejani, Johannes Totz,
  Zehan Wang, et~al.
\newblock Photo-realistic single image super-resolution using a generative
  adversarial network.
\newblock In {\em Proceedings of the IEEE conference on computer vision and
  pattern recognition}, pages 4681--4690, 2017.

\bibitem{li2018recurrent}
Xia Li, Jianlong Wu, Zhouchen Lin, Hong Liu, and Hongbin Zha.
\newblock Recurrent squeeze-and-excitation context aggregation net for single
  image deraining.
\newblock In {\em Proceedings of the European Conference on Computer Vision
  (ECCV)}, pages 254--269, 2018.

\bibitem{li2016rain}
Yu Li, Robby~T Tan, Xiaojie Guo, Jiangbo Lu, and Michael~S Brown.
\newblock Rain streak removal using layer priors.
\newblock In {\em Proceedings of the IEEE conference on computer vision and
  pattern recognition}, pages 2736--2744, 2016.

\bibitem{lim2017enhanced}
Bee Lim, Sanghyun Son, Heewon Kim, Seungjun Nah, and Kyoung Mu~Lee.
\newblock Enhanced deep residual networks for single image super-resolution.
\newblock In {\em Proceedings of the IEEE conference on computer vision and
  pattern recognition workshops}, pages 136--144, 2017.

\bibitem{loshchilov2016sgdr}
Ilya Loshchilov and Frank Hutter.
\newblock Sgdr: Stochastic gradient descent with warm restarts.
\newblock {\em arXiv preprint arXiv:1608.03983}, 2016.

\bibitem{ma2018shufflenet}
Ningning Ma, Xiangyu Zhang, Hai-Tao Zheng, and Jian Sun.
\newblock Shufflenet v2: Practical guidelines for efficient cnn architecture
  design.
\newblock In {\em Proceedings of the European conference on computer vision
  (ECCV)}, pages 116--131, 2018.

\bibitem{maas2013rectifier}
Andrew~L Maas, Awni~Y Hannun, and Andrew~Y Ng.
\newblock Rectifier nonlinearities improve neural network acoustic models.
\newblock In {\em Proc. icml}, volume~30, page~3. Citeseer, 2013.

\bibitem{nah2019ntire}
Seungjun Nah, Sungyong Baik, Seokil Hong, Gyeongsik Moon, Sanghyun Son, Radu
  Timofte, and Kyoung Mu~Lee.
\newblock Ntire 2019 challenge on video deblurring and super-resolution:
  Dataset and study.
\newblock In {\em Proceedings of the IEEE/CVF Conference on Computer Vision and
  Pattern Recognition Workshops}, pages 0--0, 2019.

\bibitem{nah2017deep}
Seungjun Nah, Tae Hyun~Kim, and Kyoung Mu~Lee.
\newblock Deep multi-scale convolutional neural network for dynamic scene
  deblurring.
\newblock In {\em Proceedings of the IEEE conference on computer vision and
  pattern recognition}, pages 3883--3891, 2017.

\bibitem{nah2021ntire}
Seungjun Nah, Sanghyun Son, Suyoung Lee, Radu Timofte, Kyoung~Mu Lee, et~al.
\newblock {NTIRE 2021} challenge on image deblurring.
\newblock In {\em IEEE/CVF Conference on Computer Vision and Pattern
  Recognition Workshops}, 2021.

\bibitem{pan2018two}
Xingang Pan, Ping Luo, Jianping Shi, and Xiaoou Tang.
\newblock Two at once: Enhancing learning and generalization capacities via
  ibn-net.
\newblock In {\em Proceedings of the European Conference on Computer Vision
  (ECCV)}, pages 464--479, 2018.

\bibitem{park2020multi}
Dongwon Park, Dong~Un Kang, Jisoo Kim, and Se~Young Chun.
\newblock Multi-temporal recurrent neural networks for progressive non-uniform
  single image deblurring with incremental temporal training.
\newblock In {\em European Conference on Computer Vision}, pages 327--343.
  Springer, 2020.

\bibitem{ren2019progressive}
Dongwei Ren, Wangmeng Zuo, Qinghua Hu, Pengfei Zhu, and Deyu Meng.
\newblock Progressive image deraining networks: A better and simpler baseline.
\newblock In {\em Proceedings of the IEEE/CVF Conference on Computer Vision and
  Pattern Recognition}, pages 3937--3946, 2019.

\bibitem{ronneberger2015u}
Olaf Ronneberger, Philipp Fischer, and Thomas Brox.
\newblock U-net: Convolutional networks for biomedical image segmentation.
\newblock In {\em International Conference on Medical image computing and
  computer-assisted intervention}, pages 234--241. Springer, 2015.

\bibitem{suin2020spatially}
Maitreya Suin, Kuldeep Purohit, and AN Rajagopalan.
\newblock Spatially-attentive patch-hierarchical network for adaptive motion
  deblurring.
\newblock In {\em Proceedings of the IEEE/CVF Conference on Computer Vision and
  Pattern Recognition}, pages 3606--3615, 2020.

\bibitem{tan2019efficientnet}
Mingxing Tan and Quoc Le.
\newblock Efficientnet: Rethinking model scaling for convolutional neural
  networks.
\newblock In {\em International Conference on Machine Learning}, pages
  6105--6114. PMLR, 2019.

\bibitem{tao2018scale}
Xin Tao, Hongyun Gao, Xiaoyong Shen, Jue Wang, and Jiaya Jia.
\newblock Scale-recurrent network for deep image deblurring.
\newblock In {\em Proceedings of the IEEE Conference on Computer Vision and
  Pattern Recognition}, pages 8174--8182, 2018.

\bibitem{tian2019fcos}
Zhi Tian, Chunhua Shen, Hao Chen, and Tong He.
\newblock Fcos: Fully convolutional one-stage object detection.
\newblock In {\em Proceedings of the IEEE/CVF International Conference on
  Computer Vision}, pages 9627--9636, 2019.

\bibitem{ulyanov2016texture}
Dmitry Ulyanov, Vadim Lebedev, Andrea Vedaldi, and Victor~S Lempitsky.
\newblock Texture networks: Feed-forward synthesis of textures and stylized
  images.
\newblock In {\em ICML}, volume~1, page~4, 2016.

\bibitem{ulyanov2016instance}
Dmitry Ulyanov, Andrea Vedaldi, and Victor Lempitsky.
\newblock Instance normalization: The missing ingredient for fast stylization.
\newblock {\em arXiv preprint arXiv:1607.08022}, 2016.

\bibitem{ulyanov2017improved}
Dmitry Ulyanov, Andrea Vedaldi, and Victor Lempitsky.
\newblock Improved texture networks: Maximizing quality and diversity in
  feed-forward stylization and texture synthesis.
\newblock In {\em Proceedings of the IEEE Conference on Computer Vision and
  Pattern Recognition}, pages 6924--6932, 2017.

\bibitem{wang2020practical}
Yuzhi Wang, Haibin Huang, Qin Xu, Jiaming Liu, Yiqun Liu, and Jue Wang.
\newblock Practical deep raw image denoising on mobile devices.
\newblock In {\em European Conference on Computer Vision}, pages 1--16.
  Springer, 2020.

\bibitem{wei2019semi}
Wei Wei, Deyu Meng, Qian Zhao, Zongben Xu, and Ying Wu.
\newblock Semi-supervised transfer learning for image rain removal.
\newblock In {\em Proceedings of the IEEE/CVF Conference on Computer Vision and
  Pattern Recognition}, pages 3877--3886, 2019.

\bibitem{whyte2012non}
Oliver Whyte, Josef Sivic, Andrew Zisserman, and Jean Ponce.
\newblock Non-uniform deblurring for shaken images.
\newblock {\em International journal of computer vision}, 98(2):168--186, 2012.

\bibitem{wu2018group}
Yuxin Wu and Kaiming He.
\newblock Group normalization.
\newblock In {\em Proceedings of the European conference on computer vision
  (ECCV)}, pages 3--19, 2018.

\bibitem{xu2013unnatural}
Li Xu, Shicheng Zheng, and Jiaya Jia.
\newblock Unnatural l0 sparse representation for natural image deblurring.
\newblock In {\em Proceedings of the IEEE conference on computer vision and
  pattern recognition}, pages 1107--1114, 2013.

\bibitem{yang2017deep}
Wenhan Yang, Robby~T Tan, Jiashi Feng, Jiaying Liu, Zongming Guo, and Shuicheng
  Yan.
\newblock Deep joint rain detection and removal from a single image.
\newblock In {\em Proceedings of the IEEE Conference on Computer Vision and
  Pattern Recognition}, pages 1357--1366, 2017.

\bibitem{yasarla2019uncertainty}
Rajeev Yasarla and Vishal~M Patel.
\newblock Uncertainty guided multi-scale residual learning-using a cycle
  spinning cnn for single image de-raining.
\newblock In {\em Proceedings of the IEEE/CVF Conference on Computer Vision and
  Pattern Recognition}, pages 8405--8414, 2019.

\bibitem{yu2018wide}
Jiahui Yu, Yuchen Fan, Jianchao Yang, Ning Xu, Zhaowen Wang, Xinchao Wang, and
  Thomas Huang.
\newblock Wide activation for efficient and accurate image super-resolution.
\newblock {\em arXiv preprint arXiv:1808.08718}, 2018.

\bibitem{yue2019variational}
Zongsheng Yue, Hongwei Yong, Qian Zhao, Lei Zhang, and Deyu Meng.
\newblock Variational denoising network: Toward blind noise modeling and
  removal.
\newblock {\em arXiv preprint arXiv:1908.11314}, 2019.

\bibitem{yue2020dual}
Zongsheng Yue, Qian Zhao, Lei Zhang, and Deyu Meng.
\newblock Dual adversarial network: Toward real-world noise removal and noise
  generation.
\newblock In {\em European Conference on Computer Vision}, pages 41--58.
  Springer, 2020.

\bibitem{Zamir2020CycleISP}
Syed~Waqas Zamir, Aditya Arora, Salman Khan, Munawar Hayat, Fahad~Shahbaz Khan,
  Ming-Hsuan Yang, and Ling Shao.
\newblock Cycleisp: Real image restoration via improved data synthesis.
\newblock In {\em CVPR}, 2020.

\bibitem{zamir2021multi}
Syed~Waqas Zamir, Aditya Arora, Salman Khan, Munawar Hayat, Fahad~Shahbaz Khan,
  Ming-Hsuan Yang, and Ling Shao.
\newblock Multi-stage progressive image restoration.
\newblock {\em arXiv preprint arXiv:2102.02808}, 2021.

\bibitem{Zamir2021MPRNet}
Syed~Waqas Zamir, Aditya Arora, Salman Khan, Munawar Hayat, Fahad~Shahbaz Khan,
  Ming-Hsuan Yang, and Ling Shao.
\newblock Multi-stage progressive image restoration.
\newblock In {\em CVPR}, 2021.

\bibitem{Zhang_2019_CVPR}
Hongguang Zhang, Yuchao Dai, Hongdong Li, and Piotr Koniusz.
\newblock Deep stacked hierarchical multi-patch network for image deblurring.
\newblock In {\em The IEEE Conference on Computer Vision and Pattern
  Recognition (CVPR)}, June 2019.

\bibitem{zhang2019deep}
Hongguang Zhang, Yuchao Dai, Hongdong Li, and Piotr Koniusz.
\newblock Deep stacked hierarchical multi-patch network for image deblurring.
\newblock In {\em Proceedings of the IEEE/CVF Conference on Computer Vision and
  Pattern Recognition}, pages 5978--5986, 2019.

\bibitem{zhang2018context}
Hang Zhang, Kristin Dana, Jianping Shi, Zhongyue Zhang, Xiaogang Wang, Ambrish
  Tyagi, and Amit Agrawal.
\newblock Context encoding for semantic segmentation.
\newblock In {\em Proceedings of the IEEE conference on Computer Vision and
  Pattern Recognition}, pages 7151--7160, 2018.

\bibitem{zhang2018density}
He Zhang and Vishal~M Patel.
\newblock Density-aware single image de-raining using a multi-stream dense
  network.
\newblock In {\em Proceedings of the IEEE conference on computer vision and
  pattern recognition}, pages 695--704, 2018.

\bibitem{zhang2019image}
He Zhang, Vishwanath Sindagi, and Vishal~M Patel.
\newblock Image de-raining using a conditional generative adversarial network.
\newblock {\em IEEE transactions on circuits and systems for video technology},
  30(11):3943--3956, 2019.

\bibitem{zhang2018dynamic}
Jiawei Zhang, Jinshan Pan, Jimmy Ren, Yibing Song, Linchao Bao, Rynson~WH Lau,
  and Ming-Hsuan Yang.
\newblock Dynamic scene deblurring using spatially variant recurrent neural
  networks.
\newblock In {\em Proceedings of the IEEE Conference on Computer Vision and
  Pattern Recognition}, pages 2521--2529, 2018.

\bibitem{zhang2020deblurring}
Kaihao Zhang, Wenhan Luo, Yiran Zhong, Lin Ma, Bjorn Stenger, Wei Liu, and
  Hongdong Li.
\newblock Deblurring by realistic blurring.
\newblock In {\em Proceedings of the IEEE/CVF Conference on Computer Vision and
  Pattern Recognition}, pages 2737--2746, 2020.

\bibitem{zhang2017beyond}
Kai Zhang, Wangmeng Zuo, Yunjin Chen, Deyu Meng, and Lei Zhang.
\newblock Beyond a gaussian denoiser: Residual learning of deep cnn for image
  denoising.
\newblock {\em IEEE transactions on image processing}, 26(7):3142--3155, 2017.

\bibitem{zhang2018shufflenet}
Xiangyu Zhang, Xinyu Zhou, Mengxiao Lin, and Jian Sun.
\newblock Shufflenet: An extremely efficient convolutional neural network for
  mobile devices.
\newblock In {\em Proceedings of the IEEE conference on computer vision and
  pattern recognition}, pages 6848--6856, 2018.

\end{thebibliography}
}

\end{document}